\documentclass[11pt]{article}

\usepackage{amsmath}
\usepackage{amssymb}

\newcommand{\beq}{\begin{equation}}
\newcommand{\eeq}{\end{equation}}
\newcommand{\bit}{\begin{itemize}}
\newcommand{\eit}{\end{itemize}}
\newcommand{\ben}{\begin{enumerate}}
\newcommand{\een}{\end{enumerate}}

\newcommand{\bs}{\boldsymbol}

\begin{document}

\begin{titlepage}

\begin{flushright}
\today
\end{flushright}

\vspace{1in}

\begin{center}

{\bf Axion-Dilaton Quantum String Cosmology for Flux Compactification and its Symmetry Breaking}

\vspace{1in}

\normalsize

{Eiji Konishi\footnote{E-mail address: konishi.eiji@s04.mbox.media.kyoto-u.ac.jp}}

\normalsize
\vspace{.5in}

 {\em Faculty of Science, Kyoto University, Kyoto 606-8502, Japan}

\end{center}

\vspace{1in}

\baselineskip=24pt
\begin{abstract}
The Wheeler-De Witt equation of type IIB quantum string cosmology compactified with constant internal $H$-fluxes on a $6$-torus, whose volume modulus is frozen, is solved under the WKB approximation. The spontaneous symmetry breaking of the S-duality group by $H$-fluxes is also examined.
\end{abstract}

\vspace{.7in}
 
\end{titlepage}

\section{{Introduction}} This paper is a supplement to a previous paper by the author and his collaborator.\cite{KM} In the previous paper, we investigated type IIB graviton-axion-dilaton string cosmology compactified on a 6-torus with an S-duality\cite{Sen,Schwarz} invariant potential for internal non-zero constant $H$-fluxes. Due to the Kaloper-Madden-Olive (KMO) no-go theorem,\cite{bv1,bv2,kmo} there is no graceful exit for the early epoch of the homogeneous Friedmann-Robertson-Walker (FRW) Universe in the presence of $H$-fluxes. That is, everywhere positive potentials of the gravi-dilaton {\it{classical}} cosmology, including the case of $H$-fluxes, do not admit a graceful exit, and the branch change from the inflation region into the FRW region always occurs on the egg-shaped region tangent to the negative region of the potential.
In the previous paper, we noted two unaddressed issues in this setup. The first issue is this graceful exit problem in {\it{quantum}} string cosmology. In the quantum regime, the birth of the expanding Universe in pre-big bang cosmology is a reflection of the wave function of the early Universe in the potential described above according to the scenario of Gasperini-Maharana-Veneziano (GMV)\cite{gmvq}. Here, we examine this issue in pre-big bang cosmology for particular $H$-fluxes by solving the Wheeler-De Witt equation\cite{WDW1,WDW2,WDW3} under the WKB approximation. The second issue is to determine the S-duality group after the spontaneous symmetry breaking induced by the axion-dilaton potential of $H$-fluxes. We examine this issue following the approach of Goldstone, Salam and Weinberg\cite{gsw}. Following the spontaneous S-duality breaking, the axion-dilaton potential of $H$-fluxes is recognized as an ``effective'' cosmological constant (in the era with the spontaneously broken S-duality symmetry) when the axion-dilaton obtain their non-zero vacuum expectation values and the volume modulus of the internal 6-torus is fixed at the Planck scale. Invoking 't Hooft's naturalness argument,\cite{tHooft} we discuss the value of the effective cosmological constant adopted by Nature according to our model.

\section{{Graceful Exit Problem in the Quantum Regime}} We consider the quantum graviton-axion-dilaton string cosmology for compactification on a $6$-torus with internal constant $H$-fluxes when the frozen volume modulus of the $6$-torus metric is Planck scale. We consider the corresponding type IIB string theory effective action with internal $H$-fluxes\cite{Schwarz,Hull}
 \begin{equation}
S=\int d^4x\sqrt{-g}\biggl[R+\frac{1}{4}{\rm{Tr}}(\partial_\mu{\cal{M}}^{-1}\partial^\mu{\cal{M}})-\frac{1}{12}{\cal{H}}_{\alpha\beta\gamma}^{(i)}{\cal{M}}_{ij}{\cal{H}}^{(j)\alpha\beta\gamma}\biggr]\;,\label{eq:startWDW}
\end{equation} in the four-dimensional FRW Einstein frame metric background. The metric $g_{\mu\nu}$ and its determinant $\sqrt{-g}$ with cosmic time $t$ and the  scale factor $a(t)$ of the Universe is given by
\begin{equation}
ds^2=-dt^2+a(t)^2\biggl(\frac{dr^2}{1-kr^2}+r^2d\Omega^2\biggr)\;,
\end{equation}where $k=0,\pm1$ is the signature of the spatial curvature, and $R$ is the space-time scalar curvature. The moduli matrix ${{\cal{M}}}$, in which the axion $\chi$ and dilaton $\Phi$ parameterize the coset $\frac{SL(2,{\bs{R}})}{SO(2)}$, and the $H$-flux vector with the internal space indices are defined by\cite{Hull}
\begin{equation}{{\cal{M}}}=\left(\begin{array}{cc}{\chi}^2
e^{{\Phi}}+e^{-{\Phi}}&{\chi}e^{{\Phi}}\\{\chi}
e^{{\Phi}}&e^{{\Phi}}\end{array}\right),\ \ \ {\cal{{H}}}_{\alpha\beta\gamma}
=\left(\begin{array}{c}{{\cal{H}}}^{(1)}\\{{\cal{H}}}^{(2)}\end{array}\right)_{\alpha\beta\gamma}\;,\end{equation}
and the rest of the background $H$-fields are set to zero.
This action Eq.(\ref{eq:startWDW}) is manifestly $SL(2,{\bs{R}})$ invariant under the transformations
\begin{equation}{{\cal{M}}}\to {\cal{O}}{{\cal{M}}}{\cal{O}}^t\;,\ \ 
{{\cal{H}}}\to({\cal{O}}^t)^{-1}{\cal{H}}\;,\ \ {g}_{\mu\nu}\to{g}_{\mu\nu}\;,
\end{equation}
for ${\cal{O}}\in SL(2,{\bs{R}})$.
We denote the final potential term in Eq.(\ref{eq:startWDW}) by $-\Lambda$. 

In the quantum cosmological scenario when $H$-fluxes are absent, we solve the corresponding Wheeler-De Witt equation in the minisuperspace of the scale factor using purely group theoretical methods by rewriting the dynamics of the axion-dilaton as the motion of a particle on the surface of a pseudosphere in the background FRW metric.\cite{Maharana} It is a well-known fact that we can decompose the ${\cal{M}}$-matrix into
\begin{equation}
{\cal{M}}=v_0{\boldsymbol{1}}+v_1\Sigma^1+v_2\Sigma^2+v_3\Sigma^3\;,\label{eq:Mdecom}
\end{equation}
for the three generators of $SL(2,{\bs{R}})$ denoted by $\Sigma^1$, $\Sigma^2$ and $\Sigma^3$ which satisfy the commutation relations
\begin{equation}
[\Sigma^1,\Sigma^2]=2\Sigma^3\;,\ \ [\Sigma^2,\Sigma^3]=-2\Sigma^1\;,\ \ [\Sigma^3,\Sigma^1]=-2\Sigma^2\;.
\end{equation}
The set of $2\times 2$ matrices satisfying the $sl(2,{\bs{R}})$ algebra are
\begin{equation}
\Sigma^1=\left(\begin{array}{cc}1&0\\0&-1\end{array}\right)\;,\ \ 
\Sigma^2=\left(\begin{array}{cc}0&1\\1&0\end{array}\right)\;,\ \ {\mbox{and}}\ \ 
\Sigma^3=\left(\begin{array}{cc}0&1\\-1&0\end{array}\right)\;.
\end{equation}
The space-time dependent coefficients of Eq.(\ref{eq:Mdecom}) are\begin{equation}v_0=\frac{\chi^2e^\Phi+e^{-\Phi}+e^\Phi}{2}\;,\ \ v_1=\frac{\chi^2e^\Phi +e^{-\Phi}-e^\Phi}{2}\;,\ \ v_2=\chi e^\Phi\;,\ \ v_3=0\;.\end{equation}
They satisfy the condition\begin{equation}v_0^2-v_1^2-v_2^2=1\;,\label{eq:relation}
\end{equation}defining the surface of a $(2+1)$-dimensional pseudosphere as mentioned. We obtain
\begin{equation}
\Lambda({\boldsymbol{v}},{\boldsymbol{w}})=\frac{v_0w_0+v_1w_1+v_2w_2}{12}\;,\label{eq:potential}
\end{equation}
where we put
\begin{eqnarray}
w_0&=&{\cal{H}}^{(1)}\cdot{\cal{H}}^{(1)}+{\cal{H}}^{(2)}\cdot{\cal{H}}^{(2)}\;,\ \ w_1={\cal{H}}^{(1)}\cdot{\cal{H}}^{(1)}-{\cal{H}}^{(2)}\cdot{\cal{H}}^{(2)}\;,\nonumber\\ w_2&=&2{\cal{H}}^{(1)}\cdot{\cal{H}}^{(2)}\;.\label{eq:fluxes}\end{eqnarray}
In Eqs.(\ref{eq:fluxes}), we denote the contraction of the internal indices of the $H$-fluxes by a dot. 
In the cosmological scenario, by rescaling the metric, redefining the matter fields accordingly and expressing the result in terms of the Einstein frame metric, the effective action Eq.(\ref{eq:startWDW}) in the minisuperspace becomes
\begin{equation}
S=\frac{1}{2}\int dt\bigl( -\dot{a}^2a+ka-a^3\eta^{ij}\dot{v}_i\dot{v}_j-a^3\Lambda({\boldsymbol{v}},{\boldsymbol{w}})\bigr)\;,\label{eq:startWDW2}
\end{equation}
where the metric in the axion-dilaton moduli space is $\eta^{ij}={\mbox{diag}}(1,-1,-1)$ with $i,j=0,1,2$. In this paper, we treat the case of $k=+1$.
The canonically conjugate momenta of the variables in Eq.(\ref{eq:startWDW2}) are
\begin{equation}
P_a=-\dot{a}a\;,\ \ P_{v_0}=-a^3\dot{v}_0\;,\ \ P_{v_1}=a^3\dot{v}_1\;,\ \ P_{v_2}=a^3\dot{v}_2\;.
\end{equation}
Then, the Hamiltonian ${\cal{H}}$ is derived from Eq. (\ref{eq:startWDW2}) and the corresponding Hamiltonian constraint is
\begin{equation}
{\cal{H}}=-\frac{1}{2}\biggl(\frac{1}{a}P_a^2+a+\frac{1}{a^3}\eta_{ij}P_{v_i}P_{v_j}-a^3\Lambda({\boldsymbol{v}},{\boldsymbol{w}})\biggr)=0\;.\label{eq:Ham}
\end{equation}
In the {\it{absence}} of the potential $\Lambda$, the action Eq.(\ref{eq:startWDW2}) and the Hamiltonian Eq.(\ref{eq:Ham}) are invariant under the $SL(2,{\bs{R}})$ transformations.
By the canonical quantization, $P_{a}\to \hat{P}_a=-i\frac{\partial}{\partial a }$ and $P_{v_i}\to \hat{P}_{v_i}=-i \frac{\partial}{\partial v_i}$, the corresponding Wheeler-De Witt equation, ${\hat{{\cal{H}}}}\Psi=0$,\cite{WDW1,WDW2,WDW3} becomes the differential equation
\begin{equation}
\biggl(\frac{\partial^2}{\partial a^2}+\frac{\partial}{\partial a}-a^2+\frac{1}{a^2}\hat{C}+a^4\Lambda\biggr)\Psi(a,{\boldsymbol{v}})=0\;,\label{eq:WDWflux}
\end{equation}
where $\hat{C}$ is the Casimir operator of $sl(2,{\bs{R}})$, which, in terms of the polar coordinate system, is just the Laplace-Beltrami operator
\begin{equation}
\hat{C}=-\frac{1}{{\rm{sinh}}\alpha}\frac{\partial }{\partial \alpha}{\rm{sinh}}\alpha\frac{\partial}{\partial \alpha}-\frac{1}{{\rm{sinh}}^2\alpha}\frac{\partial^2}{\partial \beta^2}\;.
\end{equation}
The polar coordinates are given by
\begin{equation}
v_0={\rm{cosh}}\alpha\;,\ \ v_1={\rm{sinh}}\alpha \cos\beta\;,\ \ v_2={\rm{sinh}}\alpha \sin\beta\;,
\end{equation}
with $\alpha$ is real and $0\le \beta\le 4\pi$.
 The Hamiltonian operator $\hat{{\cal{H}}}$ with $\Lambda=0$ commutes with the generators of $SL(2,{\bs{R}})$ as well as the Casimir.\cite{Maharana}
 The operator ordering of the scale factor in Eq.(\ref{eq:WDWflux}) is chosen to maintain scale factor duality.\cite{gvr}
 
We factorize the wave function of the Universe into a scale factor part and an angular part involving axion and dilaton: \begin{equation}\Psi(a,{\boldsymbol{v}})=\psi(a)Y({\boldsymbol{v}})\;.\label{eq:Psi}\end{equation}
Here we note that, although the interaction term is invariant under $SL(2,{\bs{R}})$ rotations, the axion-dilaton wave function $Y({\boldsymbol{v}})$ cannot be obtained as a representation of $SL(2,{\bs{R}})$ as was the case in the absence of $H$-fluxes.
The angular momenta for the rotations around the $v_0$, $v_1$ and $v_2$ axes are denoted by $J_0$, $J_1$ and $J_2$, respectively. Due to Eq.(\ref{eq:relation}), $J_0$ is compact and commutes with the Casimir operator of $sl(2,{\bs{R}})$.
When we choose $H$-fluxes such that \begin{equation}\Lambda({\boldsymbol{v}},{\boldsymbol{w}})=\frac{v_0w_0}{12} \;,\label{eq:adpot}\end{equation} the Hamiltonian is diagonal with respect to the Casimir operator of $sl(2,{\bs{R}})$ and $J_0$. In this case, the angular part ${Y}({\boldsymbol{v}})$ of the wave function Eq.(\ref{eq:Psi}) is in the descendent series of the representation of $sl(2,{\bs{R}})$, from its highest weight state, labeled by the eigenvalues of the Casimir operator, say $j(j+1)$, and $J_0$. Then, we can solve the scale factor part ${\psi}(a)$ of Eq.(\ref{eq:WDWflux}) approximately by using WKB methods. 
We employ the approximate ansatz for the scale factor part of Eq.(\ref{eq:WDWflux}) given by\begin{equation}
\psi(a)=A(a)e^{i{S}(a)/\hbar}\;.\end{equation}
The Wheeler-De Witt equations of $A(a)$ and $S(a)$ up to the order of $\hbar^1$ are then
\begin{subequations}
\begin{align}
-(\dot{S})^2-a^2+\frac{j(j+1)}{a^2}+a^4\Lambda&=0\;,\\
\ddot{S}+2\dot{S}\frac{\dot{A}}{A}+\dot{S}&=0\;,
\end{align}
\end{subequations}
for the orders $\hbar^0$ and $\hbar^1$ respectively.
The solution of these equations is
\begin{subequations}
\begin{align}
A(a)&={\mbox{Const}}\times \frac{e^{-\frac{a}{2}}}{\sqrt{p(a)}}\;,\\
S(a)&=\pm\int^a da^\prime p(a^\prime)+{\mbox{Const}}\;,
\end{align}
\end{subequations}
where we put
\begin{equation}
p(a)=\sqrt{-a^2+\frac{j(j+1)}{a^2}+a^4\Lambda}\;.\label{eq:p}
\end{equation}
Then, for Hubble parameter $H$, wave functions for contracting ($H<0$) and expanding ($H>0$) Universes result in the classically allowed region. (If we define a potential $V(a)$ such that $p(a)=\sqrt{-V(a)}$, then $a^2V(a)$ has two extreme values. At the smaller one ($a=0$) $a^2V(a)$ is negative. The coefficient of the $a^6$ term of $a^2V(a)$ is also negative. Thus, the sign of the potential changes zero or two times for $a>0$, dependent on the value of the axion-dilaton potential $\Lambda$. We consider the case of two sign changes and denote the value of $a$ for which the potential changes from negative to positive by $a_0$.)
The right-moving wave function $\Psi^{(+)}$ and the left-moving wave function $\Psi^{(-)}$ are given by
\begin{equation}
\Psi^{(\pm)}(a,{\boldsymbol{v}})= A_0\frac{e^{-\frac{a}{2}}}{\sqrt{p(a)}}\exp\biggl(\pm \frac{i}{\hbar}\int^a_{a_0} da^\prime p(a^\prime)\mp\frac{i\pi}{4}\biggr)Y({\boldsymbol{v}})\label{eq:wave1}
\end{equation}
and, for the region under the potential barrier,
\begin{equation}
\Psi^{(\pm)}(a,{\boldsymbol{v}})= A_0\frac{e^{-\frac{a}{2}}}{\sqrt{|p(a)|}}\exp\biggl(\pm \frac{1}{\hbar}\int^{a_0}_a da^\prime |p(a^\prime)|\biggr)Y({\boldsymbol{v}})\;.\label{eq:wave2}
\end{equation}
To show this correspondence, we use the classical mechanical relation $P_a=-a^2 H$.
The sign of $P_a$ is that of the eigenvalue of the angular part of the wave function $\psi(a)$ on the complex plane. Then, due to the existence of a positive potential, $P_a$ is positive for $\Psi^{(+)}$ and negative for $\Psi^{(-)}$.

Now we discuss the issue related to the birth of an expanding Universe in the pre-big bang scenario. It is well known that the graceful exit forbidden in classical string cosmology, due to the KMO no-go theorem, could be achieved if we adopt quantum string cosmology.\cite{gmvq}
We recall that in the pre-big bang scenario, the Universe begins in the perturbative string vacuum with negligibly small coupling and low curvature. It then enters a region of high curvature and strong coupling. In the Planckian curvature region, quantum gravity effects are dominant. When there is no potential, there is no problematic issue in the classical regime; however, this does not apply to realistic cases. When an everywhere positive potential exists, the KMO no-go theorem is then applicable in the classical regime.\cite{bv1,bv2,kmo} However, in the quantum regime, the Universe would transition to the post-big bang Universe by reflecting or anti-tunneling through classically forbidden regions of superspace with non-zero rates induced by the positive potential.\cite{gmvq} Here, we consider only the GMV reflecting case\cite{gvr}.

In quantum cosmology, one chooses suitable linear combinations of the solutions  of the Wheeler-De Witt equation, depending on the boundary conditions, to derive the wave function of the Universe. In this paper, we adopt the boundary conditions of the pre-big bang scenario. That is, we assume an initial pre-big bang configuration for the Universe and we impose the condition that in the high-curvature limit there is only a right-moving wave $\Psi^{(+)}$ approaching the singularity.\cite{gmvq}
  We denote by $\Psi_{-\infty}^{(\pm)}$ the asymptotic components of the wave function of the Universe at the limit of the shifted dilaton $\bar{\Phi}$, which takes the metric factor of the invariant volume element into the bare dilaton $\Phi$, to $-\infty$. $\Psi^{(-)}$ and $\Psi^{(+)}$ are the left-moving and right-moving parts of the wave along the space-like coordinate $\bar{\Phi}$, respectively. As $\bar{\Phi}$ tends to $-\infty$ (towards the post-big bang phase), the transition probability
\begin{equation}
R=\frac{|\Psi_{-\infty}^{(-)}(a,\bar{\Phi})|^2}{|\Psi_{-\infty}^{(+)}(a,\bar{\Phi})|^2}\;,
\end{equation}
has been shown to be non-zero for several positive dilaton potentials (applicable to the case of this section, i.e., Eqs. (\ref{eq:wave1}) and (\ref{eq:wave2}) for a positive $H$-flux potential),\cite{gmvq} even if the two branches of the classical solution are causally disconnected by a singularity.
This reflecting case and the anti-tunneling case are the scenarios for the birth of the expanding Universe in the pre-big bang solution.

\section{{S-duality Breaking by $H$-fluxes}} 
In the rest of this paper, we proceed independently from the discussions contained in previous section. We consider the spontaneously broken S-duality induced by the $H$-flux energy, which is an $SL(2,{\bs{R}})$ invariant axion-dilaton potential, and its relation to the effective cosmological constant $\Lambda$.
Throughout this section, we freeze the volume modulus of the internal 6-torus at the Planck scale. Then, the potential $-\Lambda$ in Eq.(\ref{eq:startWDW}) is an effective cosmological term by giving the vacuum expectation values to ${\cal{M}}$-matrix.
Since the following discussion does not depend on the gravitational dynamics at all, we ignore the contributions from the Einstein-Hilbert action in Eq.(\ref{eq:startWDW}).
When we switch on both $H$-fluxes, the vacuum expectation values of the axion and dilaton are the solutions of the minimality conditions for $\Lambda$:
\begin{equation}\frac{\partial \Lambda}{\partial \Phi}=0\;,\ \ \frac{\partial^2 \Lambda}{\partial \Phi^2}\ge0\;,\ \ \frac{\partial \Lambda}{\partial \chi}=0\;,\ \ \frac{\partial^2 \Lambda}{\partial \chi^2}\ge0\;,\ \ {\rm{at}}\ \Phi_0\;,\ \chi_0\;.\end{equation}
When we choose $H$-fluxes such that ${\cal{H}}^{(1)}\cdot{\cal{H}}^{(2)}=0$, the solutions are 
\begin{equation}\Phi_0={\rm{ln}}\sqrt{\frac{({\cal{H}}^{(1)})^2}{({\cal{H}}^{(2)})^2}}\;,\ \ \ \chi_0=0\;.\label{eq:sol}\end{equation}
The corresponding vacuum expectation value of the ${\cal{M}}$-matrix is
\begin{equation}{\cal{M}}_0=\left(\begin{array}{cc}\sqrt{\frac{({\cal{H}}^{(2)})^2}{({\cal{H}}^{(1)})^2}}&0\\0&\sqrt{\frac{({\cal{H}}^{(1)})^2}{({\cal{H}}^{(2)})^2}}\end{array}\right)\;.\end{equation}
When ${\cal{H}}^{(1)}\cdot{\cal{H}}^{(2)}\neq0$, we obtain a vacuum expectation value of the ${\cal{M}}$-matrix such that ${\cal{M}}_{12}$ and ${\cal{M}}_{21}$ take non-zero values.

The infinitesimal $SL(2,{\bs{R}})$ transformations are defined by
\begin{equation}
\delta^i {\cal{M}}=\bigl(1+\varepsilon \Sigma^i\bigr){\cal{M}}\bigl(1+\varepsilon (\Sigma^{i})^{t}\bigr)-{\cal{M}},\ \ i=1,2,3\;.
\end{equation}
Then the symmetry, whose transformations are generated by the generator $\Sigma^i$, is not spontaneously broken if and only if the vacuum expectation value of the ${\cal{M}}$-matrix satisfies the linear equation\cite{gsw}
\begin{equation}
\Sigma^i {\cal{M}}_0+{\cal{M}}_0(\Sigma^i)^{t}=0\;.\label{eq:SSB2}
\end{equation}
We denote the symmetric form of the ${\cal{M}}_0$-matrix by
\begin{equation}
{\cal{M}}_0=\left(\begin{array}{cc}a&c\\c&b\end{array}\right)\;.
\end{equation}
Then the conditions in Eq. (\ref{eq:SSB2}), for $i=1,2,3$, are explicitly 
\begin{equation}a=b=0\;;\ \ \ a=-b,\ c=0\;;\ \ \ a=b,\ c=0\;.
\end{equation}
Among these three conditions, only the cases with $a\neq 0$ and $b\neq 0$ have a non-trivial remaining symmetry and ${\cal{M}}_0$-matrix. However, only the $a=b$ case is possible, since the elements of the ${\cal{M}}_0$-matrix cannot have negative values.
The $a=b$ condition for the generator $\Sigma^3$ leads to
\begin{equation}
{\cal{M}}_0=\left(\begin{array}{cc}1&0\\0&1\end{array}\right)\;.
\end{equation}
This is equivalent to the following conditions on $H$-fluxes
\begin{equation}
({\cal{H}}^{(1)})^2=({\cal{H}}^{(2)})^2,\ \ \ {\cal{H}}^{(1)}\cdot{\cal{H}}^{(2)}=0\;.\label{eq:fluxcondition}
\end{equation}

The $SL(2,{\bs{R}})$ group elements for finite transformations are produced by exponentiating the $sl(2,{\bs{R}})$ generators:
\begin{eqnarray}
&&\exp(\theta\Sigma^1)=\left(\begin{array}{cc}e^{{\theta}}&0\\0&e^{-{\theta}}\end{array}\right)\;,\ \exp(\theta\Sigma^2)=\left(\begin{array}{cc}{\rm{cosh}}{\theta}&{\rm{sinh}}{\theta}\\{\rm{sinh}}{\theta}&{\rm{cosh}}{\theta}\end{array}\right)\;,
\nonumber\\&&
\exp(\theta\Sigma^3)=\left(\begin{array}{cc}\cos{\theta}&\sin{\theta}\\-\sin{\theta}&\cos{\theta}\end{array}\right)\;.
\end{eqnarray}
In our case, the remaining symmetry, which is that of the generator $\Sigma^3$, is $SO(2)$.

In the rest of this section, we discuss $H$-flux energies.
The quantized symmetry of the surviving $SO(2)$ symmetry is that generated by the transformation $g_s\to -{1}/{g_s}$ for coupling constant $g_s$. This is because the exponential map applied to $\Sigma^3$ generates the matrix $\left(\begin{array}{cc}0&1\\-1&0\end{array}\right)$ of $SL(2,{\bs{Z}})$. So, electric-magnetic duality is not broken by the $H$-fluxes. Owing to this electric-magnetic duality in type IIB string theory, we can use the well-known Dirac quantization condition on the coupling of NS-NS and R-R charges of $H$-fluxes; explicitly, $\int_{T6}{\cal{H}}^{(1)}\wedge{\cal{H}}^{(2)}$. From this and the Bianchi identity, it follows that the electric and magnetic charges of $H$-fluxes are quantized in Planck units:\cite{frev2}
\begin{equation}
\frac{1}{(2\pi\sqrt{\alpha^\prime})^2}\oint{\cal{H}}^{(1)}= m\;,\ \ \ m\in {\bs{Z}}\;,\ \ \ \frac{1}{(2\pi\sqrt{\alpha^\prime})^2}\oint {\cal{H}}^{(2)}= n\;,\ \ \ n\in {\bs{Z}}\;,\label{eq:flux}
\end{equation}
where the integrals are taken over the symplectic basis 3-cycles in $T^6$.

Since we are considering constant $H$-fluxes, the quantization conditions also hold for their squares.
Therefore, the value of the effective (positive) cosmological constant is
\begin{equation}
\Lambda_n=\frac{8}{3}\pi^4n^2(\alpha^\prime)^2\;,\ \ n\in{\boldsymbol{Z}}\;.\label{eq:spec}
\end{equation}
At this time, we have no theoretical methods to determine the cosmological constant quantitatively; however, we can deduce that the smallest value of $n$ in this spectrum Eq.(\ref{eq:spec}) is adopted by Nature.
Now, we are aware of the broken S-duality symmetry.
After the quantization of the axion-dilaton S-duality symmetry, the existence of $H$-fluxes with the conditions in Eq.(\ref{eq:fluxcondition}) spontaneously breaks the remaining $SL(2,{\bs{Z}})$ symmetry, that is, the axion shift symmetry generated by the transformation $g_s\to g_s+1$.
 The naturalness argument of 't Hooft\cite{tHooft}, which states that if the broken symmetry is restored when a parameter goes to zero then the parameter will assume a small value in the broken phase, is applied to our case by recognizing the integer $n$ as such a parameter. This naturalness argument leads to the result that Nature adopts the integer $n$ nearest to $n=0$, that is, the exact symmetry phase.\cite{Maharana1,Maharana2,Maharana3} However, as mentioned in the previous paper\cite{KM}, S-duality may be spontaneously broken in Nature to produce the vacuum expectation value of the dilaton, that is, the origin of the various coupling constants, even though it is originally an exact symmetry of string theory. So, if we assume this argument of the spontaneous breakdown of S-duality, in any context, $\Lambda_n$ for the smallest positive integer $n$ is the effective cosmological constant in the Planck units.
\section{Summary}
As a continuation of a previous paper\cite{KM} on the axion-dilaton string cosmology compactified on a $6$-torus with constant internal $H$-fluxes, we discussed issues of the quantum regime and determined the unbroken part of the S-duality group. For a specific choice of $H$-flux configurations, we partially solved the Wheeler-De Witt equation in the WKB approximation and obtained the scale factor part of the solution. We confirmed that the GMV scenario of a quantum graceful exit is valid for this case. We analyzed the part of the S-duality group unbroken by $H$-fluxes and discussed the case in which the electric-magnetic duality is kept. The $H$-flux potential is an effective cosmological constant when the volume modulus of internal torus is frozen and the vacuum expectation values of the ${\cal{M}}$-matrix are given. The volume modulus is still fixed by hand as in the previous paper\cite{KM}. We invoked the 't Hooft naturalness argument to argue for the smallness of the value of the effective cosmological constant, quantized by the Dirac condition on the $H$-flux energies, in Nature.

\section*{Acknowledgements}
I wish to thank Jnanadeva Maharana for reading the manuscript of this paper.

\end{document}